%% file: main.tex
\documentclass[sigconf,natbib=true]{acmart}

\usepackage{multirow}
\usepackage{enumitem}

\AtBeginDocument{%
  }

\copyrightyear{2026}
\acmYear{2026}
\setcopyright{cc}
\setcctype{by}
\acmConference[SIGIR '26]{Proceedings of the 49th International ACM SIGIR Conference on Research and Development in Information Retrieval}{July 20--24, 2026}{Melbourne, VIC, Australia}
\acmBooktitle{Proceedings of the 49th International ACM SIGIR Conference on Research and Development in Information Retrieval (SIGIR '26), July 20--24, 2026, Melbourne, VIC, Australia}
\acmDOI{10.1145/3805712.3808431}
\acmISBN{979-8-4007-2599-9/2026/07}

\begin{document}

\title{A General Framework for Multimodal LLM-Based Multimedia Understanding in Large-Scale Recommendation Systems}

\author{Yiming Zhu}
\authornote{Equal contribution.}
\authornotemark[2]
\affiliation{%
  \institution{Meta Platforms}
  \city{Menlo Park}
  \country{USA}
}

\author{Xu Liu}
\authornotemark[1]
\affiliation{%
  \institution{Meta Platforms}
  \city{Menlo Park}
  \country{USA}
}

\author{Ziyun Xu}
\authornotemark[1]
\affiliation{%
  \institution{Meta Platforms}
  \city{Menlo Park}
  \country{USA}
}

\author{Zheng Wu}
\affiliation{%
  \institution{Meta Platforms}
  \city{Menlo Park}
  \country{USA}
}

\author{Joena Zhang}
\affiliation{%
  \institution{Meta Platforms}
  \city{Menlo Park}
  \country{USA}
}

\author{Sirius Chen}
\affiliation{%
  \institution{Meta Platforms}
  \city{Menlo Park}
  \country{USA}
}

\author{Chenheli Hua}
\affiliation{%
  \institution{Meta Platforms}
  \city{Menlo Park}
  \country{USA}
}

\author{Silvester Yao}
\affiliation{%
  \institution{Meta Platforms}
  \city{Menlo Park}
  \country{USA}
}

\author{Qichao Que}
\affiliation{%
  \institution{Meta Platforms}
  \city{Menlo Park}
  \country{USA}
}

\author{Wentao Shi}
\authornote{Corresponding author}
\affiliation{%
  \institution{Meta Platforms}
  \city{Menlo Park}
  \country{USA}
}

\author{Junfeng Pan}
\affiliation{%
  \institution{Meta Platforms}
  \city{Menlo Park}
  \country{USA}
}

\author{Linhong Zhu}
\affiliation{%
  \institution{Meta Platforms}
  \city{Menlo Park}
  \country{USA}
}

\renewcommand{\shortauthors}{Yiming Zhu et al.}

\input{sections/abstract}



\maketitle

\input{sections/introduction}
\input{sections/related_work}
\input{sections/methodology}
\input{sections/experiments}

\input{sections/conclusion}

\section{Main Presenter Bio}
Yiming Zhu is a senior machine learning engineer at Meta, working on generative AI and recommendation. Yiming is actively working on Large Language Models and their applications. Yiming has extensive experience working on improving large-scale recommendation systems in many different stages, including end-to-end recommendation infra, model architecture, and feature engineering.

\bibliographystyle{ACM-Reference-Format}
\bibliography{refs}

\end{document}

%% file: sections/abstract.tex
\begin{abstract}
Conventional recommendation systems frequently fail to fully exploit the high-dimensional semantic signals inherent in multimedia content, thereby limiting the fidelity of user preference modeling. While Multimodal Large Language Models (MM-LLMs) offer robust mechanisms for interpreting such complex data, their integration into latency-constrained, industrial-scale architectures remains a significant challenge. To address this, we propose a generalized framework for MM-LLM-driven multimedia understanding. Our methodology employs a tripartite architecture encompassing content interpretation, representation extraction, and systematic pipeline integration, instantiated via a LLaMA2-based model that generates descriptive captions subsequently ingested as tokenized categorical features. Empirical evaluation demonstrates the efficacy of this approach, yielding a $0.35\%$ increase in offline AUC and a $0.02\%$ improvement in online metrics at scale, substantiating the practical viability of leveraging MM-LLMs to enhance large-scale recommendation performance.
\end{abstract}

%% file: sections/introduction.tex
\section{Introduction}

Modern Recommender Systems (RS) operate within increasingly heterogeneous multimedia ecosystems, where user engagement is driven by a complex co-evolution of image, video, and textual content. While traditional architectures effectively leverage structured engagement signals and explicit metadata, they often fail to encapsulate the \textit{fine-grained semantics} and latent contexts inherent in raw multimedia streams. For instance, in domains such as fashion retrieval, visual nuances convey stylistic preferences that static textual descriptions cannot fully resolve. Conventional pipelines typically relegate these signals to auxiliary metadata~\cite{visualfeature}, thereby bypassing critical opportunities to align recommendations with high-fidelity user intent.

Multimodal Large Language Models (MM-LLMs) represent a transformative paradigm for addressing this semantic gap. By synthesizing semantically dense natural language descriptions from multimodal inputs, architectures such as GPT-4V bridge the divide between raw content and high-dimensional semantic space. These models can extract granular attributes from unstructured media—transforming, for example, a user-uploaded image into a structured descriptor like \textit{``a minimalist living room with a beige sectional sofa and mid-century wooden accents''}—capturing subtle latent features that elude traditional engineering approaches.

However, the integration of MM-LLMs into production-grade RS environments presents non-trivial challenges. Established architectures, such as the Deep Learning Recommendation Model (DLRM)~\cite{naumov2019deeplearningrecommendationmodel}, are optimized for processing dense and sparse features under strict latency constraints. Mapping the generative, unstructured output of MM-LLMs into representations compatible with these high-throughput systems creates a impedance mismatch.

To address this, we propose a systematic framework to operationalize MM-LLMs within the industrial recommendation stack. Our approach is structured as a three-stage pipeline:
\begin{enumerate}[leftmargin=*]
    \item \textbf{Semantic Translation:} We leverage MM-LLMs to distill visual, textual, and acoustic signals into enriched natural language descriptors.
    \item \textbf{Representation Mapping:} We employ diverse representation learning techniques to tokenize or embed these descriptors into feature vectors compatible with downstream ranking models.
    \item \textbf{Feature Injection:} We systematically integrate these transformed representations into existing RS architectures to enhance user-item relevance modeling.
\end{enumerate}

We validate this framework by developing a LLaMA2-based MM-LLM~\cite{genai2023llama} to generate real-time captions for user-generated content, which are subsequently ingested as features into a DLRM-based stack. Our contributions are three folds:

\begin{itemize}[leftmargin=*]
    \item \textbf{A Systematic Integration Framework:} We introduce a generalized, production-ready framework that bridges the gap between the generative capabilities of MM-LLMs and the structural constraints of large-scale recommendation architectures.
    \item \textbf{Multimodal Semantic Transformation:} We develop a robust pipeline for translating raw, unstructured multimedia signals into high-fidelity semantic descriptors, enabling the capture of fine-grained user intent often lost in traditional metadata.
    \item \textbf{Large-Scale Operationalization and Validation:} We demonstrate the efficacy of our approach through a LLaMA2-based MM-LLM within a DLRM-style stack. Our empirical results confirm significant performance gains across both offline benchmarks and live online metrics.
\end{itemize}

%% file: sections/related_work.tex
\section{Related Work}
This section provides a systematic review of the evolution from traditional multimodal feature fusion to the integration of large-scale multimodal models and LLM-driven recommendation paradigms.

\noindent $\bullet$ \textbf{Multimodal RS.} With the proliferation of social networks and mobile internet, recommendation content increasingly spans multiple modalities, necessitating robust Multimodal Recommender Systems (MRS). The fundamental challenge lies in extracting heterogeneous features and fusing these signals to augment traditional ID-based representations. Conventional methodologies predominantly utilize Convolutional Neural Networks (CNNs) for visual encoding~\cite{he2016vbpr} and BERT-based transformers for textual representations~\cite{wei2023multi}. To integrate these signals, cross-attention mechanisms are widely adopted to capture inter-modal dependencies~\cite{han2022modality}. Furthermore, comprehensive frameworks such as MMRec~\cite{zhou2023mmrec} have been developed to unify signal extraction, fusion, and co-training into standardized pipelines.

\noindent $\bullet$ \textbf{MM-LLM.} Recent breakthroughs in Large Language Models (LLMs) have catalyzed advancements in multimodal learning, enabling models to align representations across text, images, video, and audio. Foundational models like CLIP~\cite{radford2021learning} and ALIGN~\cite{jia2021scaling} demonstrated the efficacy of large-scale contrastive learning for vision-language alignment. Subsequent architectures, including Flamingo~\cite{alayrac2022flamingo} and BLIP-2~\cite{li2023blip}, introduced efficient modules to interface visual encoders with frozen LLM backbones. The latest generation of MM-LLMs, such as GPT-4V~\cite{achiam2023gpt} and Gemini~\cite{team2023gemini}, further underscores the potential of instruction-tuned models for sophisticated open-ended reasoning and cross-modal generation.

\noindent $\bullet$ \textbf{LLM4Rec.} The emerging field of LLM4Rec~\cite{shehmir2025llm4rec} explores the integration of LLMs into recommendation pipelines. A primary research direction leverages LLMs as pre-trained encoders to enhance content-based representation learning. For example, RLMRec~\cite{ren2024representation} utilizes LLM reasoning to generate descriptive user and item profiles, while AlphaRec~\cite{sheng2024language} projects textual metadata representations into collaborative filtering frameworks. Parallelly, another line of research reformulates recommendation as a generative task. InstructRec~\cite{zhang2025recommendation} treats recommendation as an instruction-following problem, employing instruction-based fine-tuning for end-to-end systems. Similarly, BIGRec~\cite{bao2025bi} proposes a two-step paradigm that fine-tunes LLMs to generate semantic item tokens which are subsequently mapped to the actual item space.

%% file: sections/methodology.tex
\begin{figure}[t]
\begin{center}
\centerline{\includegraphics[width=0.98\linewidth]{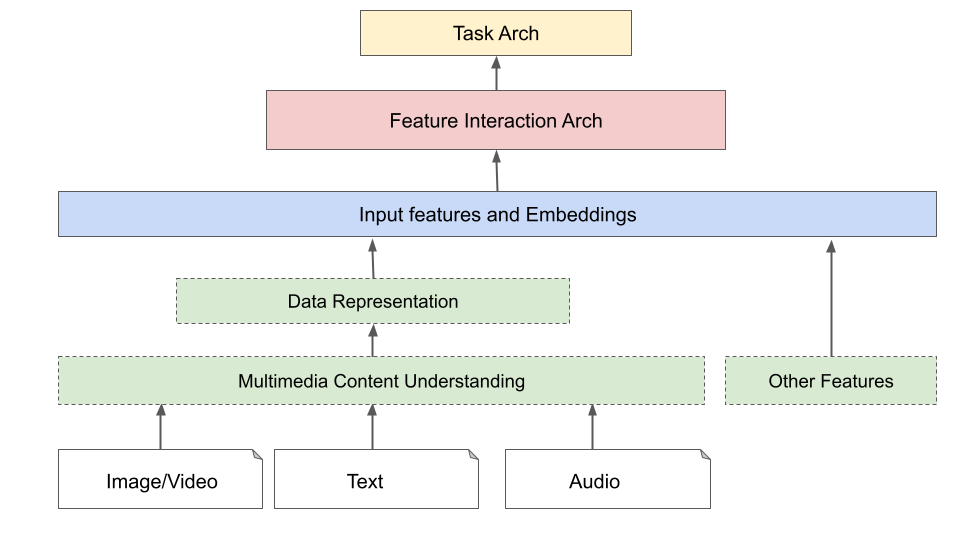}}
\caption{Overview of the Framework for MM-LLM-Based Multimedia Understanding in Recommendation.}
\label{mllm_framework}
\end{center}
\vspace{-15pt}
\end{figure}

\section{Methodology}
In this section, we introduce our framework by converting images into caption-based semantic features via an MM-LLM, then tokenizing and integrating them with standard ID embeddings to enrich ranking signals under industrial latency constraints.
\subsection{MM-LLM-Based Multimedia Understanding}

We propose a unified framework for integrating Multimodal Large Language Models (MM-LLMs) into recommender systems to enhance the semantic understanding of multimedia items. As illustrated in Figure~\ref{mllm_framework}, our architecture is predicated on three integral components, designed to bridge the gap between high-dimensional multimedia data and structured recommendation signals.

\paragraph{Multimedia Content Understanding.} 
This module utilizes an MM-LLM to extract high-fidelity semantic descriptions from raw multimedia inputs, such as imagery. Leveraging architectures capable of vision-language alignment (e.g., BLIP-style models), the system translates visual signals into natural language descriptions through strategic prompt engineering. The objective is to capture the latent semantics of the content that traditional collaborative filtering methods often fail to encode.

\paragraph{Data Representation and Alignment.} 
To render the MM-LLM outputs compatible with downstream recommendation algorithms, the generated natural language descriptions are projected into a formal feature space. This transformation encompasses processes such as tokenization~\cite{mikolov2013efficientestimationwordrepresentations} and dense embedding generation~\cite{svenstrup2017hashembeddingsefficientword}. The primary goal is to maintain semantic fidelity while mapping the unstructured text into a structured or sparse vector space suitable for ingestion by the recommendation engine.

\paragraph{Integration with Recommender Systems.} 
The final phase involves the fusion of these semantic embeddings with the recommender system's existing architecture. We achieve this by concatenating the derived MM-LLM features with traditional ID-based user and item representations. Consequently, the recommendation model is trained to optimize prediction accuracy by simultaneously leveraging collaborative signals and the enriched semantic features provided by the MM-LLM.

\begin{figure}[t]
\begin{center}
\centerline{\includegraphics[width=0.9\columnwidth]{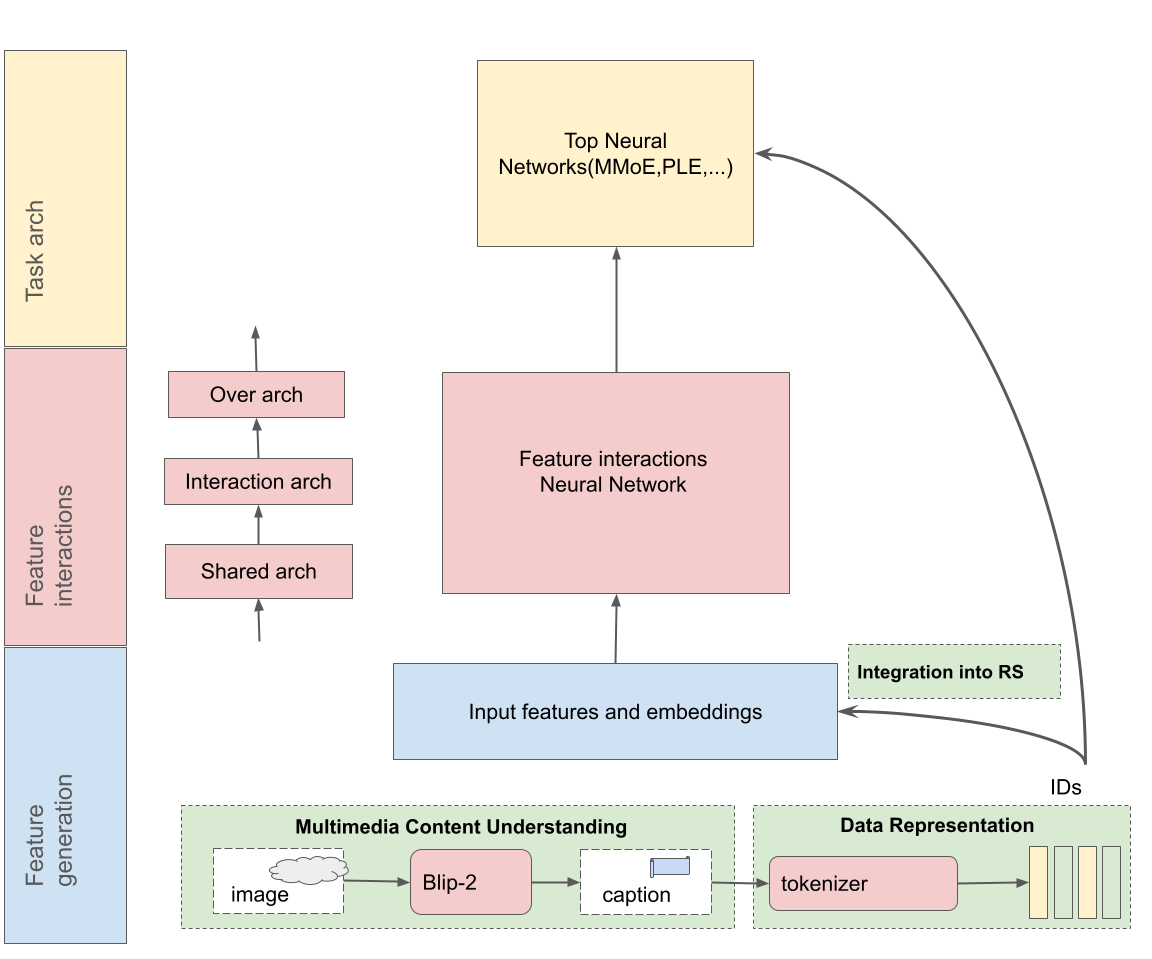}}
\caption{The structure of our approach is organized into distinct stages. The first two stages, Multimedia Content Understanding and Data Representation, are represented by the Model Encoder. In the final stage, we integrate the generated IDs into our recommendation models.} 
\label{model_approach}
\vspace{-10pt}
\end{center}

\end{figure}

\subsection{Approach}

\begin{figure}[t]
\begin{center}
\centerline{\includegraphics[width=\columnwidth]{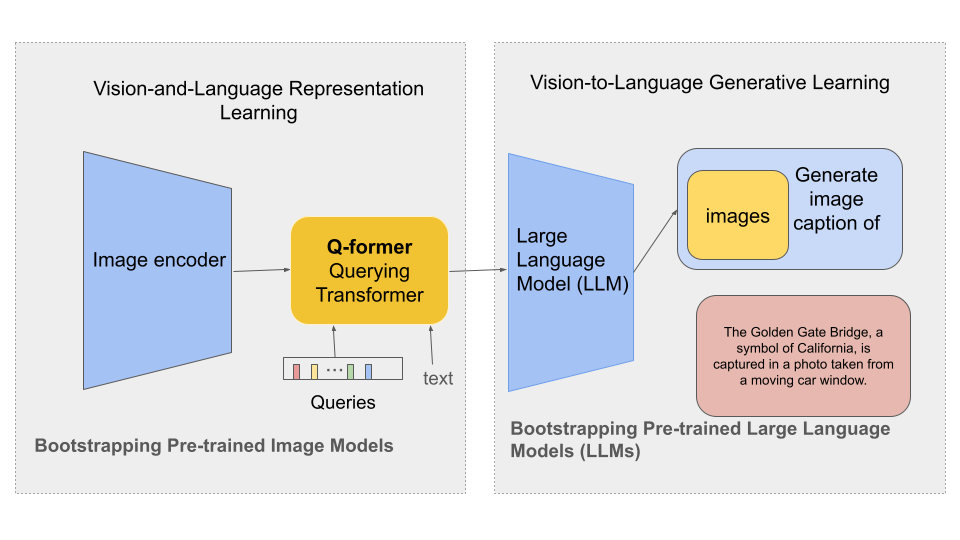}}
\caption{Overview of BLIP-2’s framework.}
\label{blip_photo}
\end{center}
\vspace{-10pt}
\end{figure}

This section details the instantiation of our framework to capture advanced visual semantics (architecture depicted in Figure~\ref{model_approach}). Current industrial recommendation models predominantly rely on static visual concepts and pre-computed embeddings. While sufficient for identifying discrete entities (e.g., \textit{`dog'}), these methods fail to encode high-order contextual dependencies (e.g., \textit{`A man playing with his dog.'}). To bridge this semantic gap in a large-scale setting, we implement our approach as follows:

\paragraph{Multimedia Content Understanding.}
The cornerstone of our framework is a Multimodal Large Language Model (MM-LLM) designed to extract dense semantic signals. Prioritizing robustness and efficiency, we adopt a two-stage architecture illustrated in Figure~\ref{blip_photo}. 

First, an image encoder processes raw visual inputs to generate robust visual embeddings. This encoder, pre-trained on massive image-text pairs, is optimized to generalize across diverse scales and resolutions. Second, we employ a generative text decoder—specifically LLaMA2~\cite{genai2023llama}—to synthesize detailed captions conditioned on these visual embeddings. To bridge the modality gap, we utilize the BLIP-2~\cite{li2023blip} paradigm, employing a lightweight Querying Transformer (Q-Former) to align visual features with the language model's embedding space via specific prompts.

To satisfy the stringent latency constraints of industrial-scale recommendation, we deploy a compact 1.5B-parameter variant of LLaMA2~\cite{genai2023llama}. This configuration ensures high Query Per Second (QPS) throughput while maintaining inference latency within strict serving budgets. Furthermore, the MM-LLM is invoked conditionally, triggering only when multimedia comprehension yields maximal marginal gain. Empirically, this design incurs negligible latency relative to the overall recommendation pipeline. By decoupled inference from the core ranking path, we ensure that system overhead and training complexity remain neutral, while significantly enhancing the semantic density of the input features.

\paragraph{Data Representation.}
To facilitate downstream learning, the raw captions undergo a normalization process to remove redundancy and ensure format uniformity. We employ a specialized tokenizer to map the generated natural language descriptions into discrete token ID sequences. Critically, these sequences are utilized not only to represent item content but also to construct user interest profiles, enabling the model to track user affinity for specific semantic concepts over time.

\paragraph{Integration into Recommender Systems.}

Finally, we integrate the MM-LLM-derived features into the recommendation architecture by concatenating tokenized content features and user interest profiles with traditional sparse ID embeddings. Feeding these enriched representations into task-specific modules enables the model to capture fine-grained interactions between visual semantics and user preferences, effectively aligning content understanding with behavioral signals.

%% file: sections/experiments.tex
\begin{table*}[t]
\caption{Experiment results of
incorporating MM-LLM features compared to the recommendation production model using visual features. Note that
+0.1\% is considered significant due to the model complexity.}
\label{exp_results}
\begin{tabular}{lccccc}
\toprule
Model & Feature Name  & Feature Description& Tokenizer Type  & AUC Gains\footnote{We use the percentage change of 1-AUC} & NE Gains \\
\midrule
Multi-task Model & MM-LLM Features & Token id lists transformed from & \multirow{2}{*} {Tokenizer A }
 &\multirow{2}{*} {0.16\%} & \multirow{2}{*} {0.07\%}  \\
 & &  generated image caption  &
 &  & \\
 \hline
\multirow{2}{*}{Multi-task Model} & MM-LLM Features  & Token id lists transformed from & \multirow{2}{*} {Tokenizer B }
 & \multirow{2}{*} {0.07\%} & \multirow{2}{*} {0.03\%} \\
 & & generated image caption  &  &
\\ \hline
\multirow{3}{*}{Multi-task Model} & MM-LLM  & Aggregated token id lists & \multirow{3}{*}{Tokenizer A}
 & \multirow{3}{*}{0.12\%} & \multirow{3}{*}{0.05\%} \\ & User Interest & based on different & &\\
& Profile Features & user engagement events & &\\

\bottomrule
\end{tabular}
\vspace{-10pt}
\end{table*}

\begin{table}[t]
\centering
\caption{Task-level offline deep dive compared to the baseline model.
Values report relative percentage reductions in NE (lower is better). 0.1\% considered significant.}
\label{tab:task_deepdive}
\begin{tabular}{l c}
\toprule
\textbf{Representative Task} & \textbf{$\Delta$NE (↓, \%)} \\
\midrule
Comment Related   & -0.11 \\
Like Related   & -0.14 \\
Share Related   & -0.07 \\
Time spent Related  & -0.04 \\
Consumption Related  & -0.14 \\
\bottomrule
\end{tabular}
\vspace{-10pt}
\end{table}

\section{Empirical Evaluations}

\subsection{Datasets}
Our evaluations are conducted on a large-scale internal dataset. The dataset consists of tens of billions of interaction examples, partitioned into training and evaluation sets at a 7:1 ratio. This setting closely mirrors the conditions faced by large-scale deployed systems.

\subsection{Results}

 To assess the effectiveness of MM-LLM generated features, we evaluate their incremental value—across various tokenizers—when integrated into a baseline recommendation model already equipped with visual features. Table \ref{exp_results} presents the offline performance on an industry-scale production model. Given the multi-task nature of the system, we report average gains in Normalized Entropy (NE) \cite{ads_at_Facebook} and AUC (measured as the percentage change of 1-AUC) across multiple tasks. To thoroughly contextualize these improvements, Table \ref{tab:task_deepdive} provides a detailed breakdown of representative tasks.

 To strengthen the reliability of these findings, we additionally conduct paired t-tests across multiple non-overlapping, independently sampled evaluation subsets of the dataset, confirming that the reported improvements are statistically significant ($p < 0.01$). We also include a comparative baseline: a recommendation production model using concatenation of visual-encoder based embeddings. As shown in Table 1, our MM-LLM features consistently outperform the baseline, indicating that the multimedia understanding capability of the MM-LLM provides unique complementary information beyond purely visual features. Furthermore, we perform an ablation study by removing the MM-LLM features from the representation. The resulting performance drop demonstrates that the observed gains are indeed attributable to the MM-LLM features.

As shown in Table \ref{exp_results}, MM-LLM features primarily drive model performance improvements. In addition, the MM-LLM user interest profile features can help the model better match user preferences and item content, thereby further improving model performance. In the stage of data representation, using different tokenizers also shows different performance. The appropriate tokenizer brings more significant improvements, which highlights the importance of tokenizers. 

In addition, we conducted a shuffle-all-based feature importance analysis, which showed that these MM-LLM features rank among the top 1\% of all production features.

We also evaluate the effect on model training and serving efficiency. Empirically, the model training QPS and the online inference cost remain neutral relative to the baseline model, demonstrating that our framework delivers accuracy gains without incurring extra latency or computational overhead.

\subsection{Online A/B Testing}
In addition to offline evaluation, we deploy the proposed features in an online experiment. The deployment yields a statistically significant +0.02\% improvement in user engagement related metrics. It further validates the industrial applicability of the proposed framework and demonstrates their effectiveness in recommendation applications.

%% file: sections/conclusion.tex
\section{Conclusion}

We present a general framework for MM-LLM-based multimedia understanding in recommendation, which abstracts the process into three core components: multimedia content understanding, data representation, and integration into RS. Our experiments demonstrate that even a preliminary instantiation of this framework can yield significant performance improvements in the large-scale recommendation application. By providing a flexible and scalable approach, this work enables the incorporation of multimedia semantics into recommendation architectures through multimodal large language model intelligence, potentially accelerating advancements in GenAI and RS.